# Neural Network-Based Detection and Multi-Class Classification of FDI Attacks in Smart Grid Home Energy Systems

**Varsha Sen[1], Biswash Basnet[2]**

Lane Department of Computer Science and Electrical Engineering, West Virginia University, Morgantown, USA

E-mail: [1] vs00039@mix.wvu.edu, [2] bb00126@mix.wvu.edu

**Abstract**

False Data Injection Attacks (FDIAs) pose a significant threat to smart grid infrastructures, particularly Home Area Networks (HANs), where real-time monitoring and control are highly adopted. Owing to the comparatively less stringent security controls and widespread availability of HANs, attackers view them as an attractive entry point to manipulate aggregated demand patterns, which can ultimately propagate and disrupt broader grid operations. These attacks undermine the integrity of smart meter data, enabling malicious actors to manipulate consumption values without activating conventional alarms, thereby creating serious vulnerabilities across both residential and utility-scale infrastructures.

This paper presents a machine learning-based framework for both the detection and classification of FDIAs using residential energy data. A real-time detection is provided by the lightweight Artificial Neural Network (ANN), which works by using the most vital features of energy consumption, cost, and time context. For the classification of different attack types, a Bidirectional LSTM is trained to recognize normal, trapezoidal, and sigmoid attack shapes through learning sequential dependencies in the data. A synthetic time-series dataset was generated to emulate realistic household behaviour. Experimental results demonstrate that the proposed models are effective in identifying and classifying FDIAs, offering a scalable solution for enhancing grid resilience at the edge. This work contributes toward building intelligent, data-driven defence mechanisms that strengthen smart grid cybersecurity from residential endpoints.

## 1. Introduction

The modernization of power systems into smart grids has introduced a wide range of advantages through smart metering, automation, and two-way communication between consumers and energy providers. Integration at the residential level of smart meters and Home Area Networks (HANs) has supported real-time energy monitoring, dynamic tariffs, and better demand management. However, this enhanced digitalization and utilization of communication protocols have made these systems vulnerable to cyber threats. Cyber-attacks on advanced power systems have become a critical concern, particularly following the first-of-its-kind cyber-attack on Ukraine's power grid on December 23, 2015[1]. In response to such attacks, considerable efforts have been channelled toward detecting, preventing, and mitigating industrial cyberattacks, including time synchronization attacks, replay attacks, denial-of-service (DoS) attacks, man-in-the-middle (MITM) attacks, and false data injection attacks (FDI) [2],[3]. These attacks compromise energy data's integrity, potentially leading to incorrect billing, false estimations of load, and broader system-level instabilities.

There is a large body of literature available on detecting and mitigating FDIAs in smart grid networks using machine learning and model-based techniques [4]. These works presume extensive synchronized data from Phasor Measurement Units (PMUs) and concentrate on securing centralized state estimation processes. In contrast, very few works address FDIA detection in decentralized and resource-constrained environments such as Home Area Networks (HANs) [5]. HANs, consisting of smart meters, appliances, and local controllers, are especially vulnerable due to limited computational capacity, lack of supervision, and direct end-user access to devices [6]. With the increasing data-driven residential systems, addressing FDIA threats in HANs is crucial to ensure end-to-end grid security.

In addition to technical disruption, FDIAs create regulatory, financial, and privacy concerns through user-level device targeting and exploitation of weak authentication protocols in distributed networks such as Home Area Networks (HANs) [7]. Moreover, attackers may exploit temporal patterns or economic signals—such as time-of-use pricing—to design context-aware attacks coinciding with consumption peaks or DER scheduling patterns. To resolve this, the utilization of machine learning techniques has gained prominence, allowing data-driven models to identify anomalous patterns that deviate from normal grid behaviour [8]. Recent works on fault location and network reconfiguration in distribution and transmission systems have similarly emphasized the need for structural adaptability and fault-resilient topology management, reinforcing the urgency for real-time intelligent solutions in smart grid security [9], [10].

Furthermore, attackers design dynamic, shaped injections—such as trapezoidal or sigmoid profiles—that mimic natural load variations and reactive power compensation dynamics, and therefore are difficult to distinguish from legitimate fluctuations [11],[12]. Identifying not only the presence but also the nature of FDIA is essential to facilitate adaptive countermeasures and perform targeted forensic analysis. Sparse like and evasive FDIA that affect only subsets of system measurements but can induce catastrophic errors if not detected in time. Consequently, there is a growing need for architectures that support distributed intelligence, adaptive learning, and lightweight anomaly classification—especially for HAN gateways, micro-PMUs, and residential inverters [13].

Given these needs, we present a dual-stage framework that first detects FDIAs using an Artificial Neural Network (ANN) and then classifies attack patterns (normal, trapezoidal, sigmoid) with a Bidirectional Long Short-Term Memory (BiLSTM) network. The models are trained on synthetically generated yet behaviorally accurate residential data, incorporating temporal and cost-based features, offering a practical solution for real-time HAN-level cyber-resilience.

## 1.1 False Data Injection Attack

False Data Injection Attacks (FDIAs) are a very malicious and destructive class of cyberattacks posing serious threats to the integrity of smart grid infrastructure, inflicting severe consequences. The main aim of FDIAs is to compromise the accuracy of measurement data essential for real-time monitoring, state estimation, and regulation in cyber-physical power systems. Unlike traditional anomalies, FDIAs are carefully crafted to evade traditional Bad Data Detection (BDD) mechanisms by systematically injecting forged but believable data into the system. This sophisticated tactic allows attackers to manipulate the operating dynamics of the grid without triggering alarm systems, thus creating opportunities for financial gain, privacy loss, or even large-scale power disruptions [8].

Within Home Area Networks (HANs), False Data Injection Attacks (FDIAs) have the potential to compromise both network-level and device-level components. At the device level, attackers can use vulnerable smart meters to create or manipulate energy readings, leading to erroneous reporting and consumption pattern distortion. At a network level, attackers can capture and manipulate communication traffic using tactics such as the Man-in-the-Middle (MITM) attack, which makes it possible for attackers to inject forged measurements into the sensor network directly. This can include data tampering, distortion, or injection of false signals, greatly threatening the trustworthiness and transparency of the grid [2], [11].

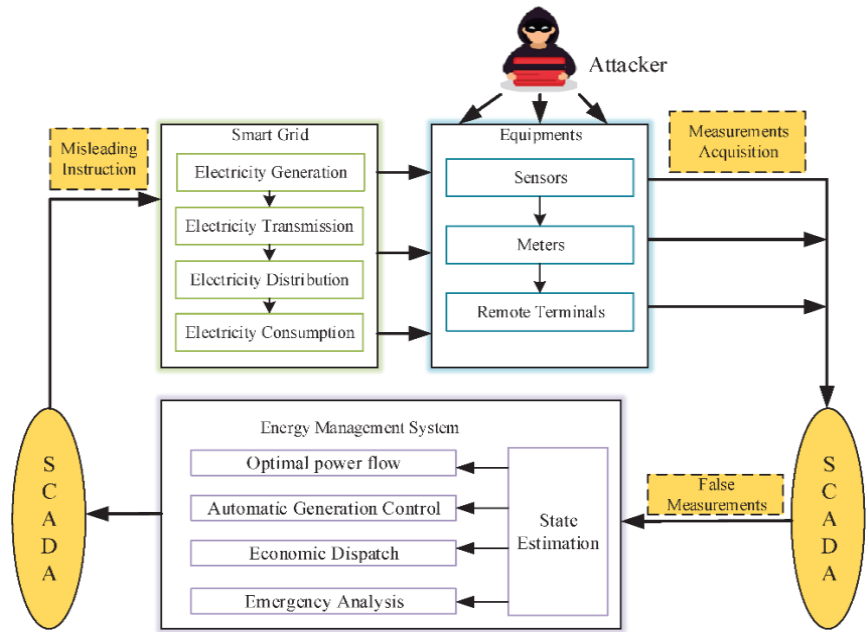

Figure 1: FDI Attack in Smart Grid [15]

Outside the residential domain, undetected FDIAs originating from HANs can propagate to distribution and transmission networks, which leads to incorrect system state estimations, destabilized grid operations, financial losses, and potential blackouts. By distorting state variables, FDIAs may lead to false control and destabilize power flow. Attackers can manipulate pricing mechanisms by targeting specific buses, influencing locational marginal prices (LMPs) and real-time market signals [16]. Studies have shown that compromised estimations can disrupt economic dispatch algorithms and alter optimal power generation schedules [17]. In more advanced cases, attackers may exploit these vulnerabilities for financial gain by launching sustained FDIAs that manipulate bidding behaviors and market settlements. Research in [18] has demonstrated that virtual bidding strategies enabled by false data injection can result in continuous financial gains, ultimately undermining market fairness and stability. The growing sophistication of FDIAs, combined with the decentralization of the smart grid, highlights the urgent need for intelligent, data-driven detection mechanisms capable of operating effectively in real-time HAN environments.

## 2. Related Work

Recent smart grid cybersecurity developments have focused intensely on the identification and classification of False Data Injection Attacks (FDIAs) due to their capability to compromise metering data without triggering conventional alarms like bad data detection (BDD) mechanisms. Early research [19] demonstrated that blind attacks can be crafted with the knowledge of the measurement Jacobian matrix, which renders state estimators ineffective. To alleviate this, several works explored data-driven approaches, namely the use of machine learning (ML) for attack detection and classification.

Support Vector Machines (SVM) and K-Nearest Neighbours (KNN) have most frequently been utilized for binary detection of FDIAs. In [20], a comparison between supervised and unsupervised models showed that SVMs can detect FDIAs from statistical deviations in measurement vectors. These models are, nevertheless, restricted in scalability and do not model temporal dependencies.

To go beyond such limitations, Artificial Neural Networks (ANNs) have been employed to learn nonlinear feature relationships with more flexibility to noisy and high-dimensional data [21]. Further enhancement has been achieved by deep learning architectures such as Recurrent Neural Networks (RNNs) and Long Short-Term Memory (LSTM) networks. Such architectures exploit temporal sequences in the load and voltage measurements for detecting stealthy attacks with high sensitivity and specificity [22], [23]

Recent studies have also been on federated learning [24], semi-supervised methods [25], and graph-based architectures [26] to support distributed environments and reduce the reliance on labelled data. Federated deep learning architectures have specifically been proposed for privacy-conscious FDIA detection on decentralized smart grid assets. In another direction, heuristic feature selection techniques such as Genetic Algorithms (GA), Particle Swarm Optimization (PSO), and Binary Cuckoo Search (BCS) have been used to reduce computational complexity without compromising classification performance [12].

On the classification side, multiclass models have been suggested for classifying different attack strategies. For instance, [27] proposed a multi-label deep learning model for categorizing different kinds of FDIA, while [28] used predicted residuals from load estimators for anomaly classification.

## 3. Methodology

This section outlines the data generation process and the two-stage machine learning framework employed for False Data Injection (FDI) detection and classification in residential smart grid data.

### A. Synthetic Data Generation

To develop and evaluate supervised machine learning models for False Data Injection (FDI) detection and classification in Home Area Networks (HANs), we generate a high-resolution synthetic dataset that simulates both normal and attack conditions. The residential electricity demand profile is modelled using a bottom-up approach inspired by Muratori et al. [29], which captures occupant behaviour, appliance usage, and HVAC operations at 10-minute intervals over a full calendar year. This realistic modelling enables fine-grained emulation of household energy usage variability across time.

The significance of using two distinct attack types—trapezoidal and sigmoid—is to expose AI models to varied and realistic adversarial patterns. The trapezoidal attack mimics sharp, peak-hour falsifications typically used to maximize economic disruption. This includes a flat-top region, time-aligned with high-tariff hours, and a superimposed ripple to simulate camouflaged but malicious activity. On the other hand, the sigmoid attack models a gradual, long-duration deviation in load, common in stealthy cyber-physical intrusions. These attacks increase slowly over time and include engineered spikes to emulate spoofed events such as unauthorized electric vehicle charging.

In our simulation, the year-long load profile is divided into three temporal blocks: six months of normal data, three months of data modified with trapezoidal attacks, and three months with sigmoid attacks. Electricity prices are modelled using a time-of-use structure with higher rates during peak hours. Each data record is labelled in two formats to support different learning tasks. For multi-class classification, we assign: 0 for normal, 1 for trapezoidal, and 2 for sigmoid attacks. For binary detection, both attack types are grouped as class 1 (attack), and normal data is labelled 0. This dual labelling ensures that the dataset is compatible with both binary anomaly detectors and multi-class classifiers.

### B. ANN Architecture for FDI attack detection model

An Artificial Neural Network (ANN) is implemented to detect False Data Injection (FDI) attacks in residential smart grid data. The network architecture includes four input neurons corresponding to energy consumption, cost, hour of the day, and day of the month. These attributes are holistically combined with an input layer of 100 neurons, followed by one output neuron for binary classification: normal (0) or FDI attack (1). The artificial neural network (ANN) is realized using the Multilayer Perceptron (MLP Classifier) in the scikit-learn package, and the data is split in the ratio 80/20 for training and testing, respectively.

The hidden layer uses the activation function Rectified Linear Unit (ReLU), which can be formulated as $g(z) = \max(0, z)$, where z is the weighted input [22]. The use of ReLU maintains large gradients for positive inputs, which promotes quicker and steadier convergence; this property is particularly suitable in detecting slight fluctuations in energy consumption patterns. The training process involves two main stages: forward propagation and backward propagation, enabling the network to learn complex consumption behavior and detect FDI anomalies.

a. Forward propagation: It is the process through which input features pass through different layers of the network, each calculating activations over weighted inputs that are summed through activation functions. The output neuron then produces a prediction, a3, which is tested against the actual label y using the Mean Squared Error (MSE) loss function, given as:

$$\text{Loss} = 0.5*(y-a_3)^2$$

This loss quantifies the deviation between predicted and actual classes, where minimizing enhances the ANN's ability to distinguish between genuine and falsified readings. The forward propagation process is illustrated in Figure. 2.

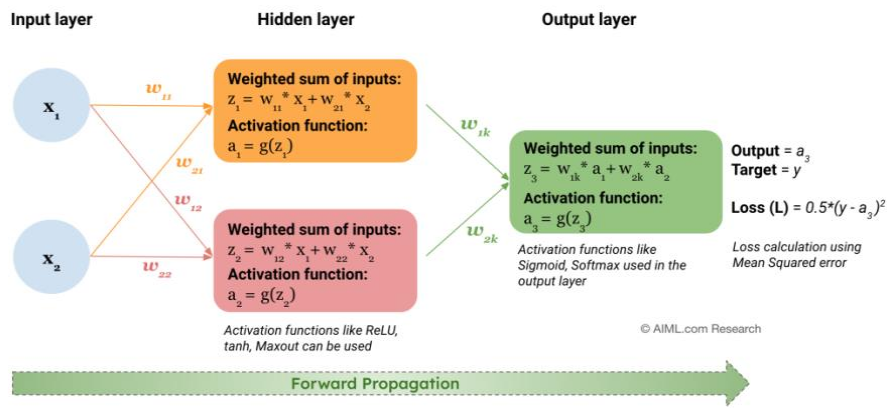

**Figure 2.** ANN Forward Propagation [30]

b.  Backward propagation: Backward propagation calculates the gradient of the loss function in terms of the weights using the chain rule. This provides information that is utilized to update the weights to minimize the error, following the Stochastic Gradient Descent (SGD) algorithm, which is augmented by the Adam optimizer. This iterative process allows the network to detect FDI attacks in terms of anomalies in consumption during peak and off-peak hours. A graphical explanation of backward propagation in terms of artificial neural networks is depicted in Figure 3.

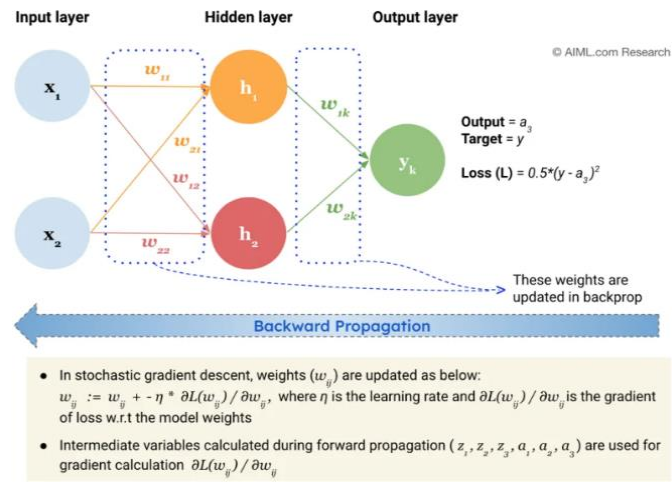

**Figure 3.**  ANN Backward Propagation [30]

To evaluate robustness, the model was trained on datasets of varying sizes and tested under several activation functions, namely ReLU, Sigmoid, and Tanh. As seen in Figure 6, ReLU exhibits consistently better accuracy and better generalization towards real-time fault detection and isolation (FDI).

**C. LSTM Architecture for FDI Attack Classification Model**

The Bidirectional LSTM architecture was selected for FDI attack classification in smart home energy systems for its ability to learn bidirectional temporal dependencies in sequential energy consumption patterns, which is especially critical for distinguishing between shaped FDI attack profiles like trapezoidal and sigmoid curves. The model was trained on a synthetically generated high-resolution dataset comprising three classes—normal, trapezoidal, and sigmoid attacks.

Feature engineering included time-based features (hour, day), cyclical transformations (sin/cos of hour and day), the energy-to-cost ratio, and a peak hour indicator. All features were scaled using min–max normalization to ensure scale invariance. To analyze comprehensively the temporal dynamics in customer usage patterns, sequences consisting of 8 overlapping time steps were generated to predict the class label for the next time step. The resulting model consisted of two Bidirectional LSTM layers stacked on top of one another to learn past and future relationships in time, dense and dropout layers, and one softmax output layer to classify among three classes. The model was trained using categorical cross-entropy loss to generalize well and the AdamW optimizer to ensure convergence stability. Early stopping and model checkpointing routines were applied to prevent overfitting. Model performance metrics consisted of overall accuracy, confusion matrix, and per-class precision and recall. Figure 4 depicts the entire architecture.

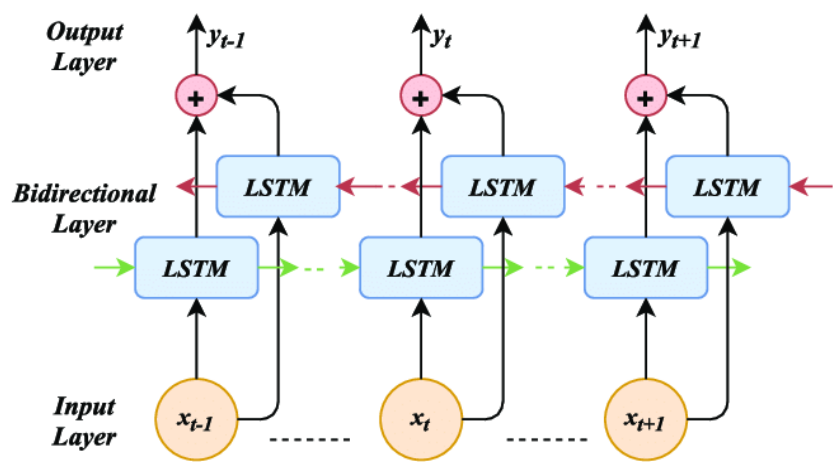

**Figure 4.** Bidirectional LSTM Architecture[31]

4. **Results And Discussion**

An Artificial Neural Network (ANN) with ReLU activation was developed to detect False Data Injection (FDI) attacks in residential smart grid data. The model uses four input features: energy consumption, cost, hour, and day, selected for their relevance to typical usage behaviour, availability, and sensitivity to anomalies.

The ANN model for FDIA detection achieved an accuracy of 97.68%, confirming its strong ability to distinguish between normal and attack conditions. The model's performance was further validated using the Mean Squared Error (MSE) loss metric, reflecting minimal deviation between predicted and actual labels. With a single hidden layer of 100 neurons, the model balances simplicity and high performance, which makes it ideal for real-time deployment in resource-constrained Home Area Networks (HANs). The evaluation metrics are defined mathematically as follows in Table 1.

$$Accuracy = \frac{Number\ of\ classified\ Events}{Total\ Events}$$

**Table 1.** Fault detection results

| Metric | Value (%) |
|---|---|
| **Accuracy** | 97.68 |
| **MSE** | 0.0232 |

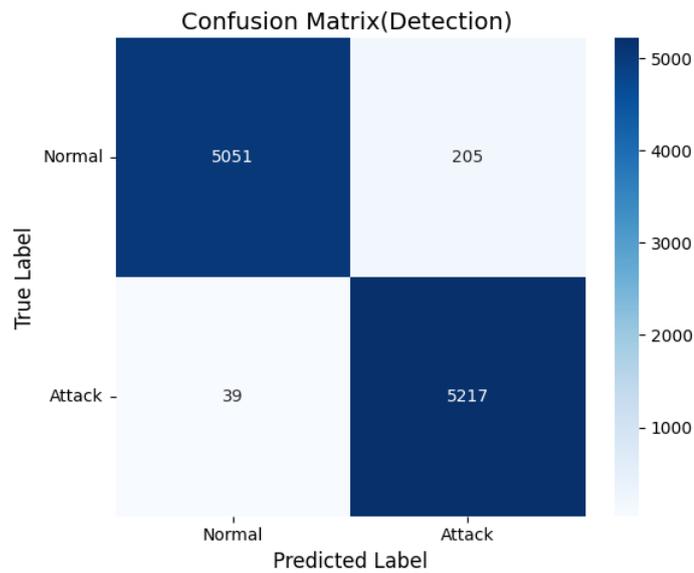

**Figure 5.** FDI Attack Detection Confusion Matrix

As illustrated in Figure. 6, the ReLU-based ANN consistently outperformed Sigmoid and Tanh activation functions across increasing dataset sizes. ReLU Activation function surpasses 96% accuracy on training sets of 50000 data, while the alternatives plateaued below 92%. This confirms ReLU's superior generalization and learning efficiency.

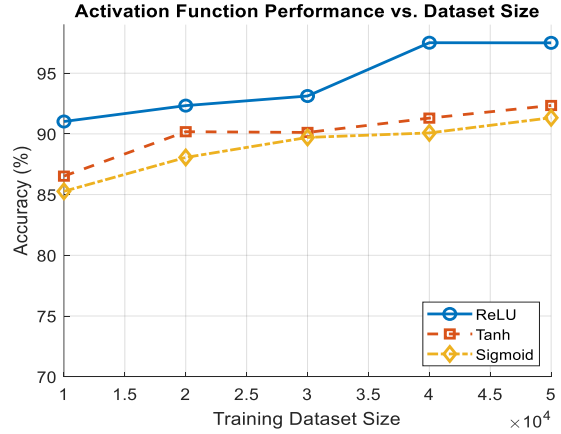

**Figure 6.** Impact of activation function on detection accuracy across varying training dataset sizes

The proposed Bidirectional LSTM model classifies three types of data categories: normal operation, trapezoid FDI attacks, and sigmoid FDI attacks. The training process of the model exhibited convergence at 67 epochs due to the usage of early stopping and checkpointing techniques, which were validation accuracy-dependent. The model exhibited 90.88% precision on the test data, showing its strong generalization power to unseen data. In addition, the discriminative performance of the model, shown in Figure 7, is represented in the confusion matrix.

The evaluation metrics are defined mathematically as follows in Table 2.

**Table 2.** Classification Results

| Metric | Value (%) |
|---|---|
| **Accuracy** | 90.88 |
| **MSE** | 0.046677 |

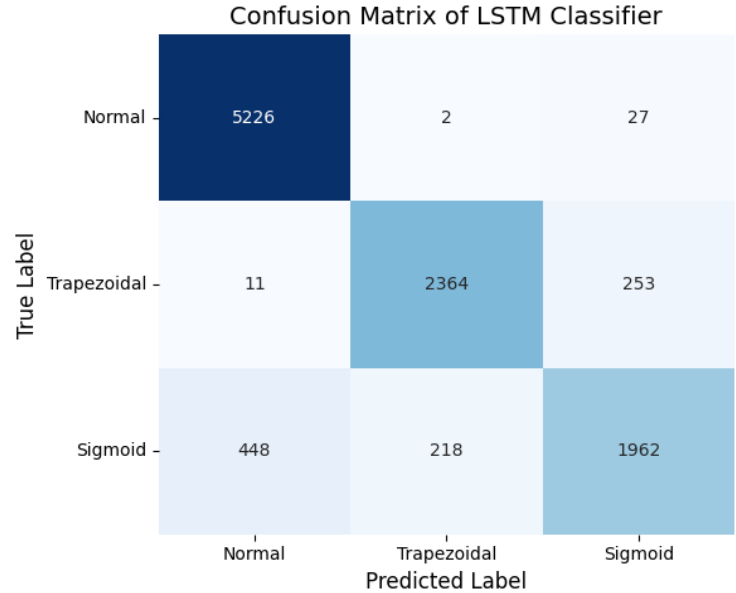

**Figure 7.** FDI Attack Classification Confusion Matrix Using LSTM

As shown, the model performs exceptionally well on normal class detection, with minimal false positives. For trapezoidal attacks, the classifier maintains high precision, misclassifying only 264 instances (8.2%) as either normal or sigmoid. Notably, sigmoid attacks are more challenging to detect, with ~21.4% misclassified, mainly as normal. This reflects the nature of sigmoid attacks—designed to mimic gradual load variations and evade abrupt anomaly detection thresholds.

Despite this, the model provides better performance for all classes, a feature particularly useful for FDI attack defence mechanisms that call for differentiation among multiple types of attacks to apply effective countermeasures. Adding label smoothing and bidirectional temporal modelling dramatically improves the model's robustness towards noisy or uncertain data instances. Together, these results validate the effectiveness of using sequence-based deep learning models, like BiLSTM, in comprehensive FDI attack classification in smart grid Home Area Networks (HANs) utilizing energy data from smart meters.

## 5. Conclusion

This paper introduces an intelligent FDI attack detection and classification scheme in smart grid Home Area Networks using machine learning. A lightweight ANN made it possible to detect attacks in a speedy way using key behavioral features, while a Bidirectional LSTM network found usage in learning temporal patterns in multi-class classifications. Choosing a Bidirectional LSTM comes from the fact that it can effectively capture both past and future temporal dependencies in energy-consumption sequences, essential for modeling the rise–

plateau–decline behavior of trapezoidal and sigmoid FDI attacks, at the same time mitigating vanishing gradient issues common in standard RNNs. Techniques suggested in this paper exhibited strong discriminative and generalization ability in diverse attack scenarios.

While the results validate the effectiveness of the approach, there is still scope for development and improvement of accuracy. Some of the areas for future research include utilizing attention mechanisms to achieve more temporal sensitivity, using ensemble methods for more robustness, and adding real-time streaming for actual smart grid deployment in real-world environments. Domain adaptation and transfer learning can also promise more generalized performance across different grid scenarios and customer loads.